\documentclass[preprint,aps,preprint,amsmath,amssymb]{revtex4-1}
\usepackage{graphicx}
\pdfoutput=1
\usepackage{amsmath}
\usepackage{color}
\usepackage{float}
\setcitestyle{super}
\usepackage[english]{babel}
\usepackage{color,soul}

\begin{document}
\title{Superconductivity in silicon: a review}

\author{Monika Moun}
\author{Goutam Sheet}
\email{goutam@iisermohali.ac.in}

\affiliation{Department of Physical Sciences, Indian Institute of Science Education and Research (IISER) Mohali, Sector 81, S. A. S. Nagar, Manauli, PO: 140306, India.}

\begin{abstract}

\textbf{Silicon, one of the most abundant elements found on Earth, has been an excellent choice of the semiconductor industry for ages. Despite it's remarkable applications in modern semiconductor-based electronic devices, the potential of cubic silicon in superconducting electronics remained a challenge because even heavily doped silicon crystals do not superconduct under normal conditions. It is apparent that if superconductivity can be introduced in cubic silicon, that will bring a breakthrough in low-dissipation electronic circuitry. Motivated by this, attempts have been made by several research groups to induce superconductivity in silicon through a number of different routes. Some of the other structural phases of silicon like $\beta$-Sn and simple hexagonal are, however, known to display superconductivity. In the present review article, various theoretical and experimental aspects of superconductivity in silicon are discussed. Superconductivity in different phases and different structural forms of silicon are also reviewed. We also highlight the potential of superconducting phases of silicon for technological applications in super-conducting nano-electronics.}

\end{abstract}


\maketitle

\section*{I. Introduction}

Silicon has been a material of choice for semiconductor technology for the ultimate control that has been achieved on electrical properties.\cite{WN, Kim, Ieong, Duv} Furthermore, silicon displays outstanding thermal stability and is abundantly found on Earth. Functionality of silicon can be further enhanced by incorporating other novel physical properties such as magnetism and superconductivity.\cite{Jansen, Delmo, si10, Cava} The electrical conductivity of crystalline silicon could be greatly altered in a control way by doping with both electrons and holes, a dissipationless superconducting state has been a far reaching property at reasonable temperature\cite{Kittel, Dai} Theoretical investigation suggested that superconductivity can be obtained in doped elemental semiconductors like silicon and germanium under certain conditions.\cite{si8, Conne, Boeri}  Following such studies, experimentalists all over the world have been trying to investigate a superconducting phase in silicon. It was seen that superconductivity could possibly appear in silicon under very high pressure.\cite{Pressure_Si} The cubic diamond phase of silicon has also manifested superconductivity at extreme conditions i.e. low temperatures and heavy doping beyond the solubility limit.\cite{Anton, si11} Recently, a superconducting phase in moderately doped crystalline silicon was studied under non-superconducting metallic point contacts depicting a high critical temperature of 11 K.\cite{Silicon_APL} Besides the bulk form of silicon, it's different structural forms including silicene and nano-structured silicon have shown potential for superconducting applications with relatively higher transition temperatures. This review article provides an overview of theoretical and experimental investigations in the emerging field of superconductivity in different phases of silicon. The review article is structured as follows: various phases of silicon including bulk superconducting phases, superconductivity in doped silicon and it's different structural forms will be discussed. Superconducting phase under point-contacts in doped silicon will also be reviewed followed by the author's perspective and potential applications of superconductivity in silicon.

\section*{II. Various phases of silicon}

The pioneering work of Minomura and Drickamer discussed various pressure induced phase transitions in different semiconductors including silicon, GaAs, GaSb, InP, InAs, AlSb and germanium.\cite{Minomura} For all the semiconductors, a phase transition to a metallic state was observed at high applied pressures of the order of a few GPa. In silicon, the metallic phase was obtained at applied pressure of nearly 19-20 GPa. While the cubic diamond phase of silicon has been the most commonly investigated of all structures, it's high-pressure phases have also been examined both experimentally and theoretically.\cite{H, Hu, SJ} The phase diagram of silicon manifests several metastable phases also under the ambient conditions (Figure 1). Under applied pressure of upto $\sim$ 11 GPa, the cubic diamond (Si-I) structure of bulk silicon is preserved.\cite{Mujica} It undergoes phase change from cubic diamond phase (Si-I) to body centered tetragonal ($\beta$-Sn, Si-II) at applied pressure of slightly above 11 GPa. The $\beta$-Sn(Si-II) phase changes to an orthorhombic Imma phase and transforms to a simple hexagonal (sh, Si-V) at a pressure of about 15 GPa. Structural phase change from simple hexagonal phase (sh) to a hexagonal closed packed (hcp) occurs at a pressure around 40 GPa with an intermediate orthorhombic Cmca phase (Si-VI) as revealed from x-ray diffraction studies and ab-initio calculations.\cite{Stefan} A face centered cubic (fcc) phase of silicon is obtained at a pressure of 79 GPa, which remains stable upto the highest investigated pressure of 248 GPa.\cite{Duclos} 
\begin{figure}[h!]
	
	\includegraphics[width=1.0\textwidth]{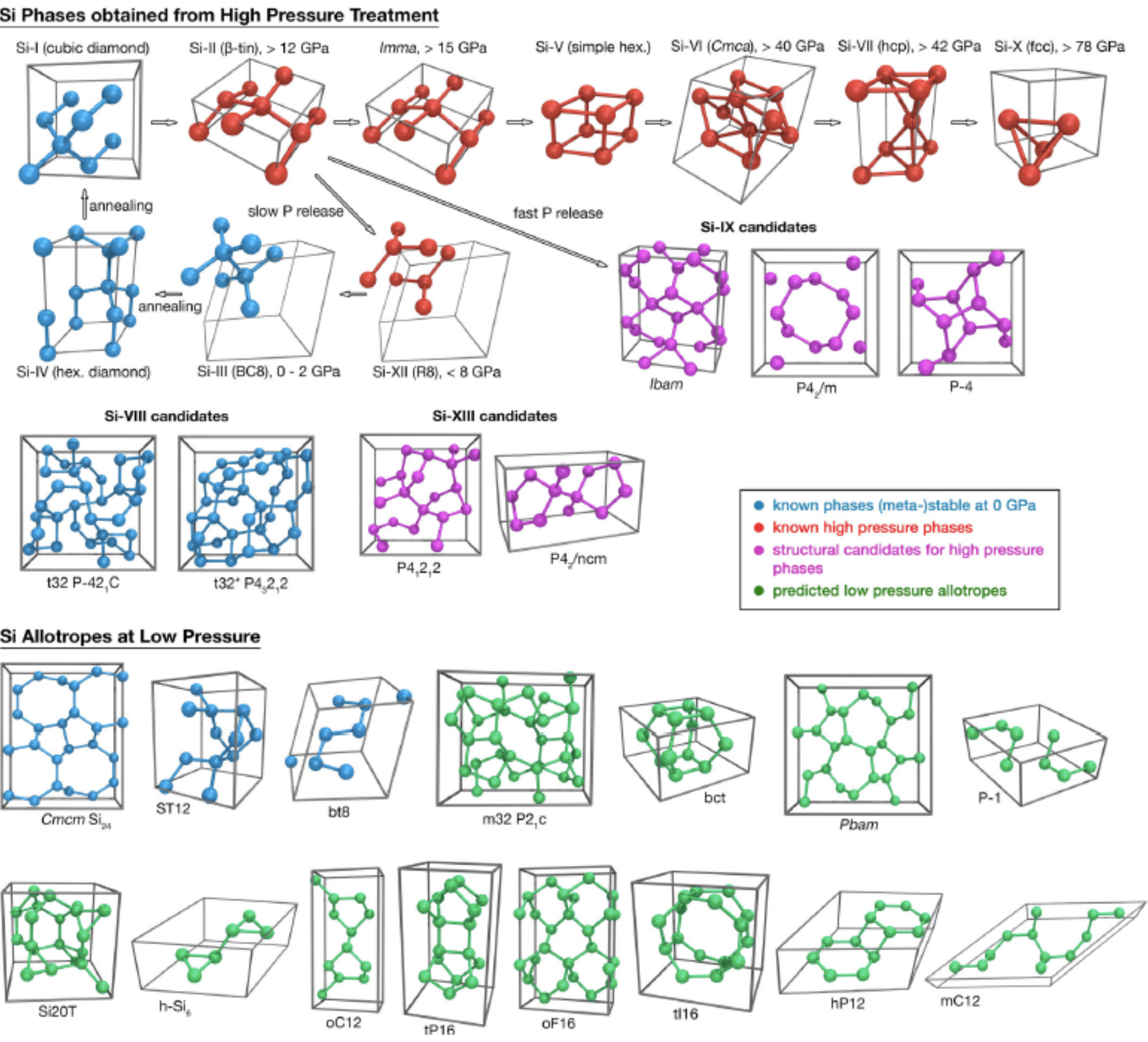}
	\caption{Schematic representation of various low and high pressure structural phases of silicon\cite{Stefan}, reproduced with permission, 2016, American Institute of Physics}
	\label{s1}
\end{figure}
Interestingly, all the phases of silicon above pressure of nearly 12 GPa are metallic in nature. This semiconductor-metal transition between cubic diamond phase to high pressure metallic phase is irreversible. Apart from the stable high-pressure silicon phases, a series of metastable phases has also been produced as a result of releasing pressure instead of recovering the cubic diamond phase under different conditions.
When pressure is slowly released from $\beta$-Sn(Si-II) phase, a metastable body centered cubic phase bc8 (Si-III) has been observed with an intermediate phase called rhombohedral (r8) phase.\cite{Wentorf} Two other metastable phases of silicon Si-VIII and Si-IX consisting of large tetragonal cells have been obtained under rapid decompression. Si-IX phase was examined to be the mixture of two structures having space groups P42/m and P-4.\cite{Nguyen} Two more silicon phases were theoretically predicted, ST12 phase with distorted tetragonal cells\cite{Clark} and a body centered tetragonal (bct) phase\cite{Fujimoto}. Figure 1 depicts various low and high pressure structural phases of silicon.

\section*{III. Bulk superconducting phases of silicon}

In this section, we discuss the high pressure structural phases of silicon in which superconductivity was observed. The theoretical findings of structural, vibrational and electronic properties of silicon at high-pressure performed by Chang and Cohen indicated that a pressure-sensitive soft-transverse-acoustic phonon mode could result in the successful phase-transition from $\beta$-Sn(Si-II) to simple hexagonal (sh) to hexagonal closed packed (hcp) structure.\cite{Chang} The high density of states at the Fermi level of simple hexagonal (sh) phase of pressurized silicon, and strong covalent bonding in the transverse direction might lead to local field effects which could improve the electron-electron interaction in sh silicon. This may facilitate the appearance of a superconducting phase in the sh phase of silicon. The observation of superconductivity in sh phase of Si was also suggested by Needs and Martin from self-consistent density functional theory.\cite{Needs} Experimentally, the resistance vs temperature ($R-T$) measurements down to liquid helium temperatures were done by Wittig on $\beta$-Sn(Si-II) and a discontinuity in the resistance-temperature curve was reported at 6.7 K.\cite{Wittig} This was attributed to a superconducting transition in the $\beta$-Sn(Si-II) phase of bulk silicon. Il’ina and Istkevich observed a similar discontinuity at 7 K for $\beta$-Sn(Si-II).\cite{G} Following these studies, superconductivity was also theoretically verified and experimentally examined for $\beta$-Sn(Si-II) and sh metallic phases of silicon at low temperatures and high pressures by Chang and Cohen.\cite{Cohen} For the $\beta$-Sn(Si-II) phase, the measured critical temperature was 6.3 K at a pressure of 12 GPa which was lower than the previous reports. Critical temperature was observed to go to maximum of 8.2 K at a pressure of 15.2 GPa for the sh phase of silicon. 

\section*{IV. Superconductivity in boron doped silicon}

In 2006, the discovery of superconductivity in doped cubic phase of silicon by Bustarret \textit{et al.} opened a new era of research on such superconducting transition in doped silicon based systems.\cite{si11} Until this discovery, superconductivity had only been obtained in high-pressure structural phases of silicon. Bustarret \textit{et al.} showed that substitutional doping of boron achieved using gas immersion laser doping technique beyond it's equilibrium solubility in silicon could induce superconductivity. Critical temperature and critical field for the boron-doped (B concentration: 6 x 10$^{20}$) silicon was found to be 0.35 K and 0.4 T, respectively as measured by electrical resistivity and magnetic susceptibility measurements. \textit{Ab initio} calculations were performed to calculate electron-phonon coupling, superconducting transition temperature and the microscopic pairing mechanism in B-doped silicon. The electron-phonon coupling constant was found to be 0.28 which well matched with the previous studies on pressurised silicon. The study also  suggested that the observed transition in silicon could be accounted by standard BCS mechanism and even higher critical temperature could be possible via high boron doping in silicon. 

Following this experimental observation, Bourgeois \textit{et al.} investigated the vibrational and electronic modes along with electron- phonon coupling in cubic phase of silicon with varying doping elements using a first-principles approach.\cite{E} Their investigation provides evidences that the as-discovered superconductiivity in highly doped silicon could be explicated on the basis of a standard phonon-mediated BCS mechanism. In addition, authors suggested a possible way to increase the critical temperature by one order of magnitude via incorporating aluminium doping in silicon rather than boron.  

\section*{V. Superconductivity in silicene}
The low-dimensional systems for applications in nanoscale superconducting devices have come up recently as a possible alternative with the possibility of higher $T_{c}$.\cite{Blase} In fact, 2D materials can be ideal candidates for such application as it has been predicted theoretically that in highly doped 2D materials, critical temperature could go higher than the boiling point of liquid nitrogen.\cite{Savini} Silicene is the monolayer counterpart of silicon having honeycomb lattice structure like graphene and has proved it's potential to be compatible with current nano-technology. 
Several semiconductors from group IV have been found to be superconductors upon heavy boron doping. In these strong covalent materials such as silicon and diamond, charge doping occurs into $sp^{3}$-bonding bands near the Fermi level having large electron-phonon coupling.\cite{KW} Considering the case of 2D materials, for instance, pristine graphene has $sp^{2}$ bonding but hole-doped graphene favors $sp^{3}$ bonding and thus, enhanced electron-phonon coupling.\cite{Savini} Silicene can exhibit even stronger electron-phonon coupling owing to low-buckled honeycomb structure which may lead to partial $sp^{3}$ hybridization.\cite{Wan} In that case, silicene may also superconduct.

\begin{figure}[h!]
	
	\includegraphics[width=1.0\textwidth]{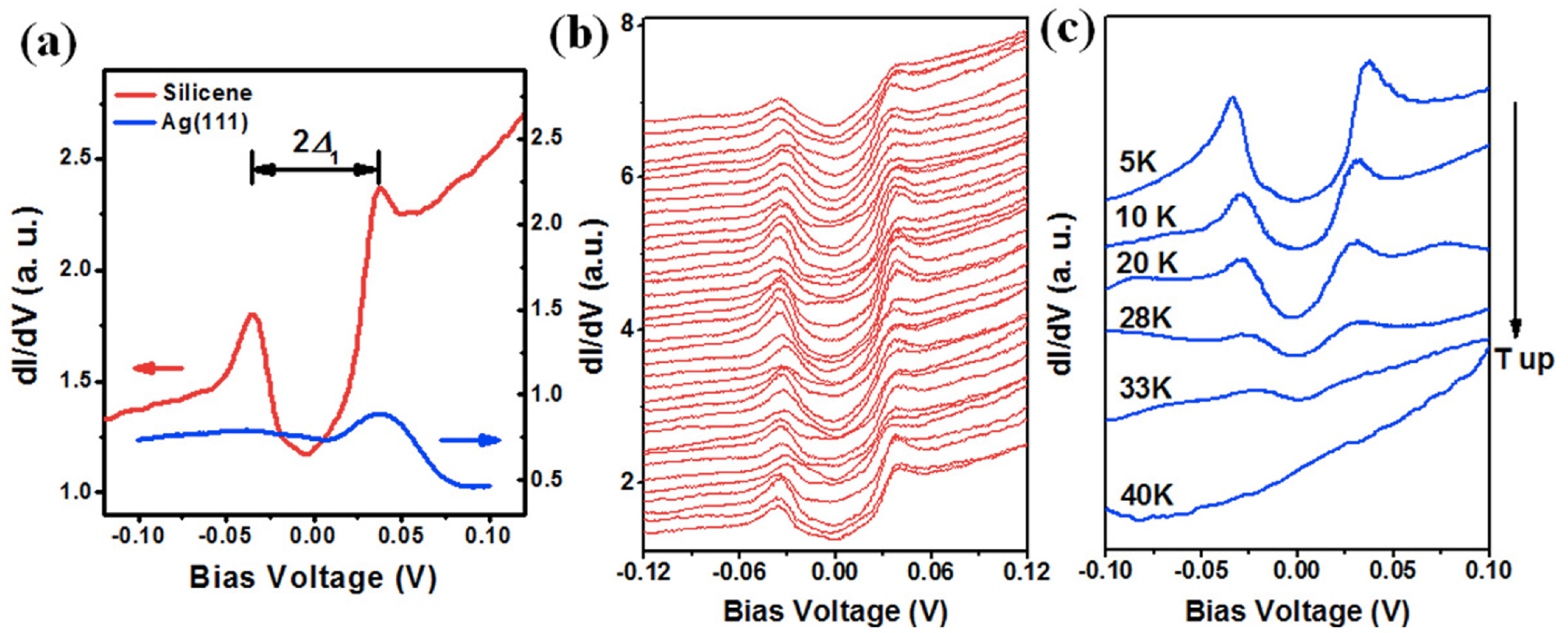}
	\caption{(a) Conductance spectra acquired on silicene and silver substrate (Ag (111)), respectively. (b)  Conductance spectra curves taken along silicene film at various positions, exhibiting the uniformity of the superconducting gap. (c) Temperature dependent conductance spectra on silicene film.\cite{Ag}, reproduced with permission, 2013, American Institute of Physics}
	\label{s1}
\end{figure}

Experiments show that the charge carriers in silicene on Ag (111) behave as massless Dirac Fermions.\cite{Feng} Following this study, a superconducting gap was observed experimentally using scanning tunneling spectroscopy in silicene on Ag(111) surface with the relatively higher $T_{c}$ values ranging from 35 – 40 K.\cite{Ag} The superconductor-metal transition was investigated by temperature-dependent transport and tunneling experiments shown in figure 2. The gap observed in spectrum appeared precisely at the Fermi energy strongly hinting to the possibility of the gap being a superconducting gap. The amplitude of the gap was measured to be 35 meV. It was proposed that the charge transfer between Ag and silicene along with strong electron-phonon coupling led to the high $T_{C}$ superconductivity in silicene/Ag. However, the knowledge of the superconducting pairing mechanism is still lacking, hence the mechanism behind superconductivity in silicene remains an open problem. Following the experimental observations, Wan\textit{ et al.} theoretically studied the superconductivity in silicene upon application of strain using first principles calculations.\cite{Wan} The group suggested that electron-doped silicene could serve as a good 2D superconductor with increased density of states under biaxial tensile strain and the critical temperature could go higher above 10 K by 5 \% applied strain. The enhancement in the transition temperature was attributed to the strong interaction between out of plane phonon modes transverse to the silicene plane and additional strain induced electronic states near the Fermi level.

\section*{VI. Superconductivity in nanostructured silicon}

$p$-type silicon quantum wells on $n$-type silicon have also shown potential towards enhanced superconducting properties in silicon. Bagraev \textit{et al.} fabricated heavily boron doped $p$-type silicon quantum wells on the $n$-type Si(100) with high mobility.\cite{Spin, SC} The group proposed that the formation of oxide overlayers on silicon surfaces could promote the generation of the self-interstitial sites and vacancies known as 
``microdefects". Longitudinal and lateral quantum wells between microdefect layers were formed depending on the oxide layer thickness. Although both longitudinal and lateral silicon quantum wells could be used as electrical and optical active microcavities for nanoelectronic and optoelectronic applications, the dangling bonds present at the interfaces could hinder the practical applications. Therefore, short-time boron diffusion has been suggested for the passivation of silicon surface as well as the created defects during oxidation process of the silicon wafers. This aids the transformation of the microdefect arrays into $\delta$-barriers as observed from the infrared Fourier spectroscopy, cyclotron resonance and scanning tunneling microscopy (STM) techniques. Possible superconducting gap of 0.022 eV and high superconducting transition temperature of 150 K was observed.\cite{Si NS} 
\begin{figure}[h!]
	
	\includegraphics[width=1.0\textwidth]{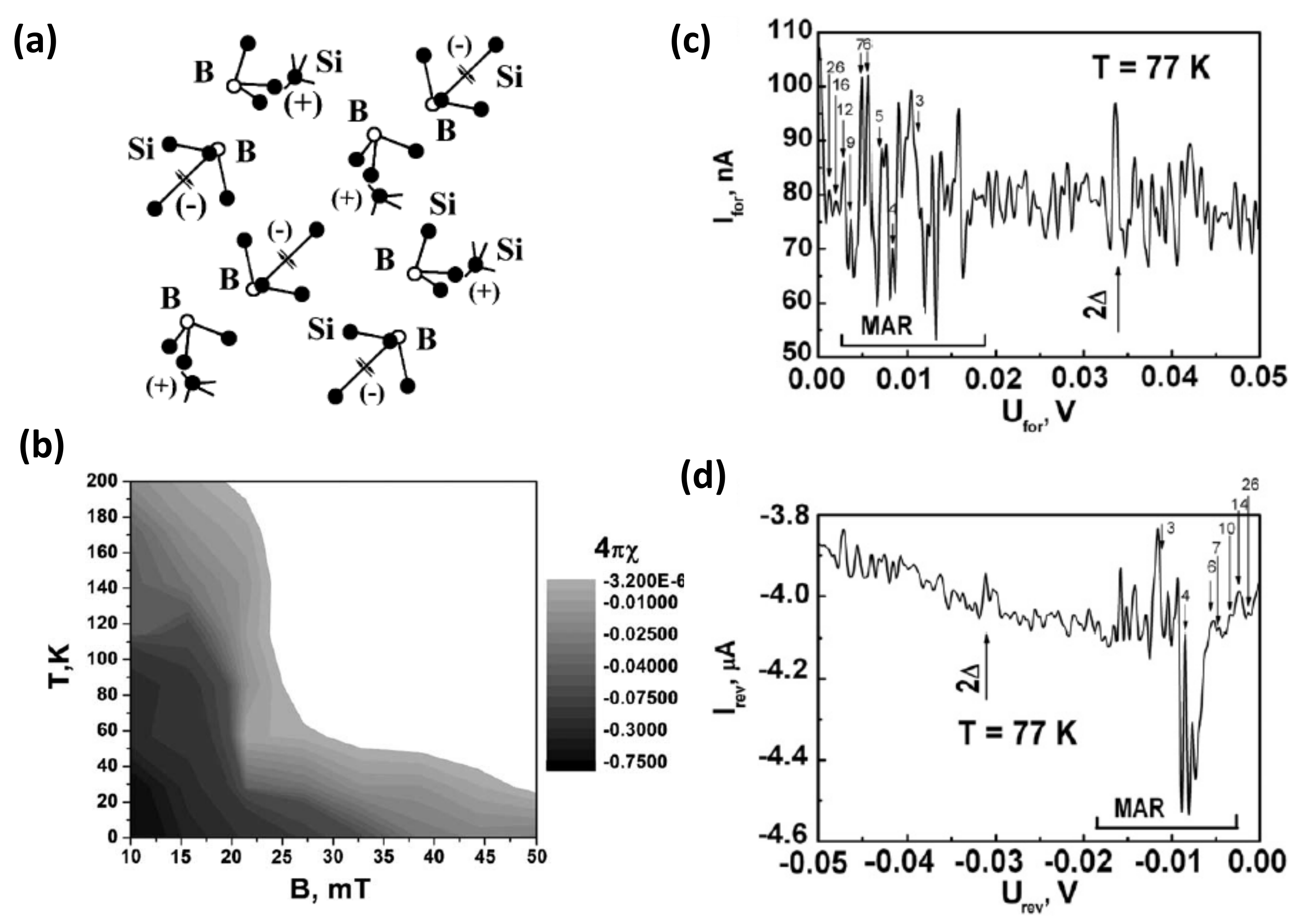}
	\caption{(a) Proposed mechanism for the boron dipole centers (b) Dependence of temperature and magnetic field on static magnetic susceptibility in $\delta$ barriers confined silicon quantum wells.\cite{Si NS}, reproduced with permission, 2006, Elsevier (c-d) Multiple Andreev reflection (MAR) signatures with the (c) forward bias and (d) reverse bias applied to the silicon quantum wells\cite{W}, reproduced with permission, 2008, Elsevier}
	\label{s1}
\end{figure}
The same group investigated the detailed hole transport in the silicon sandwich nanostructures using tunneling spectroscopy.\cite{LE} They showed that the multiple Andreev reflection features of holes in the silicon quantum wells could provide the mechanism behind the proximity effect phenomenon in silicon sandwich nanostructures. Moreover, the oscillations of current-voltage characteristics at temperatures above and below $T_{c}$ were identical reflecting the presence of interaction between the single hole tunneling and Cooper pairs. Their observation could lead to controlled fabrication of nano-dimensional Josephson junctions and hybrid systems suggesting new possibilities in the field of superconducting nanoelectronics.\cite{CWJ, PJ}

\section*{VII. Superconductivity under point contacts on heavily doped silicon}
Recently, evidence of superconductivity was found in moderately doped silicon crystals under point contacts with non-superconducting metallic tips with reasonably high critical temperature($\sim11 K$).\cite{Silicon_APL} 
\begin{figure}[h!]
	
	\includegraphics[width=1.0\textwidth]{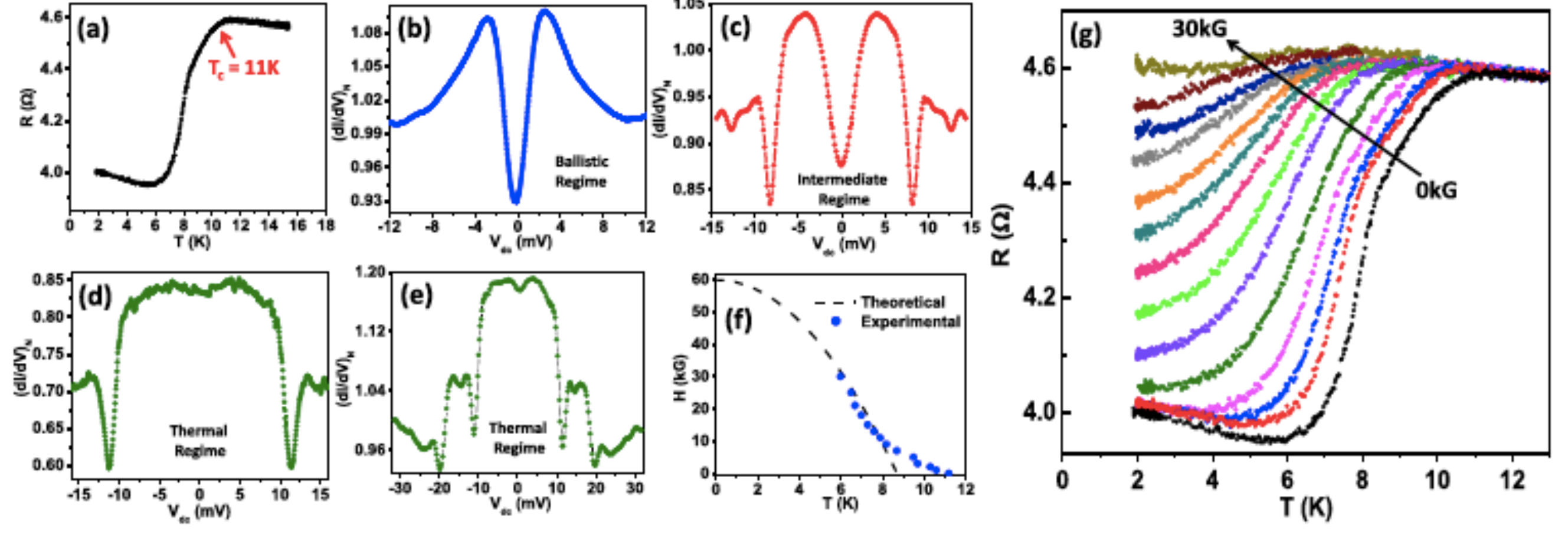}
	\caption{(a) $R-T$ plot depicting superconducting transition for point contacts in intermediate regime shown in (c). Conductance spectra obtained on As-doped Si(111) showing (b) ballistic regime (c) intermediate regime (d) thermal transport regime  (e) conductance spectra for thermal regime depicting multiple critical current peak features. (g) $R -T$ curves at different magnetic fields (f) $H – T$ phase diagram obtained from (g).\cite{Silicon_APL}, reprinted with permission, 2018, American Institute of Physics}
	\label{s1}
\end{figure}
The superconducting phase was examined in As-doped Si (111) using point contact spectroscopy and different transport regimes i.e ballistic, intermediate and thermal regime were investigated. The comparative estimate of the contact diameter with the mean free path of electrons decide the transport regimes.\cite{BTK} Ballistic transport regime is said to occur in the condition when the contact diameter becomes smaller than electron's elastic mean free path. On the other hand, in thermal regime, the contact diameter is larger than the electron's inelastic mean free path. For a point-contact between metallic tip and the sample, according to the Wexler's formula, total resistance is sum of both thermal or Maxwell's resistance $(R_{M} = \rho(T)/2r)$ and ballistic or Sharvin's resistance $(R_{S} = \frac{2h/e^{2}}{{(rk_{F})}^{2}})$ where $\rho(T)$, r, $k_{F}$, h and e represent the resistivity of the sample, radius of the tip, Fermi radius, Planck's constant and electronic charge, respectively. In the intermediate regime, the contact diameter is such that both ballistic resistance and thermal resistance become comparable. The conductance spectra reveals different signatures for each kind of transport. If the conductance spectra ($dI/dV$ vs $V$) shows symmetric two peaks about V=0, it is the signature of Andreev reflection, known to appear in ballistic regime of superconducting point contacts.\cite{GoutamPRB} Next comes the "intermediate regime" where the conductance spectrum displays distinct dips accompanying suppressed Andreev reflection dominated symmetric double peak feature. In the thermal regime, strong conductance dips are observed with no signature of Andreev reflection associated conductance peaks. The spectroscopic data shown in figure 5 confirms the presence of a superconducting phase in doped-crystalline silicon under point contacts. The superconducting phase was obtained with a considerable high $T_{c}$ 11 K and superconducting energy gap of 2 meV.\cite{Silicon_APL} shown in figure 4.
Subsequently, the superconducting phase under point contacts was examined on both $n$-doped (As dopants) and $p$-doped (B-dopants) facets of silicon crystals using various metallic tips exhibiting universal behavior of mesoscopic superconductivity in silicon crystals.\cite{Moun} \\
Before this, point-contact spectroscopy was been used to understand semiconductor-metal junctions by several groups. Tanaka \textit{et al.} suggested that superconductivity in silicon could be obtained in rapidly quenched granules of silicon without applying high pressure.\cite{Tanaka} Conductance spectra was acquired in both Au-Si and Si-Si point contact junctions. It was proposed that the observation of symmetric negative  $dI/dV$ spike features at non-zero voltages was due to the growing quenched metallic phase granules (figure 5). These granules led to lattice condition similar to the high pressure superconducting phase of silicon and possibly created a microjunction network in the vicinity of point contact. This study followed the similar investigations performed in different materials such as GaAs, PbSe etc. In the model as proposed by Shaw, the sample fabrication process of point contacts and $p-n$ junctions involves heat treatment which may create a micro-Josephson junction's network.\cite{Shaw} These consist of superconducting granules in the boundary of the tunnel junction, with the condition of presence of superconducting element in the materials comprising of the junction (for example, Pb in PbSe junction, Ga in GaP). 
\begin{figure}[h!]
	
	\includegraphics[width=0.8\textwidth]{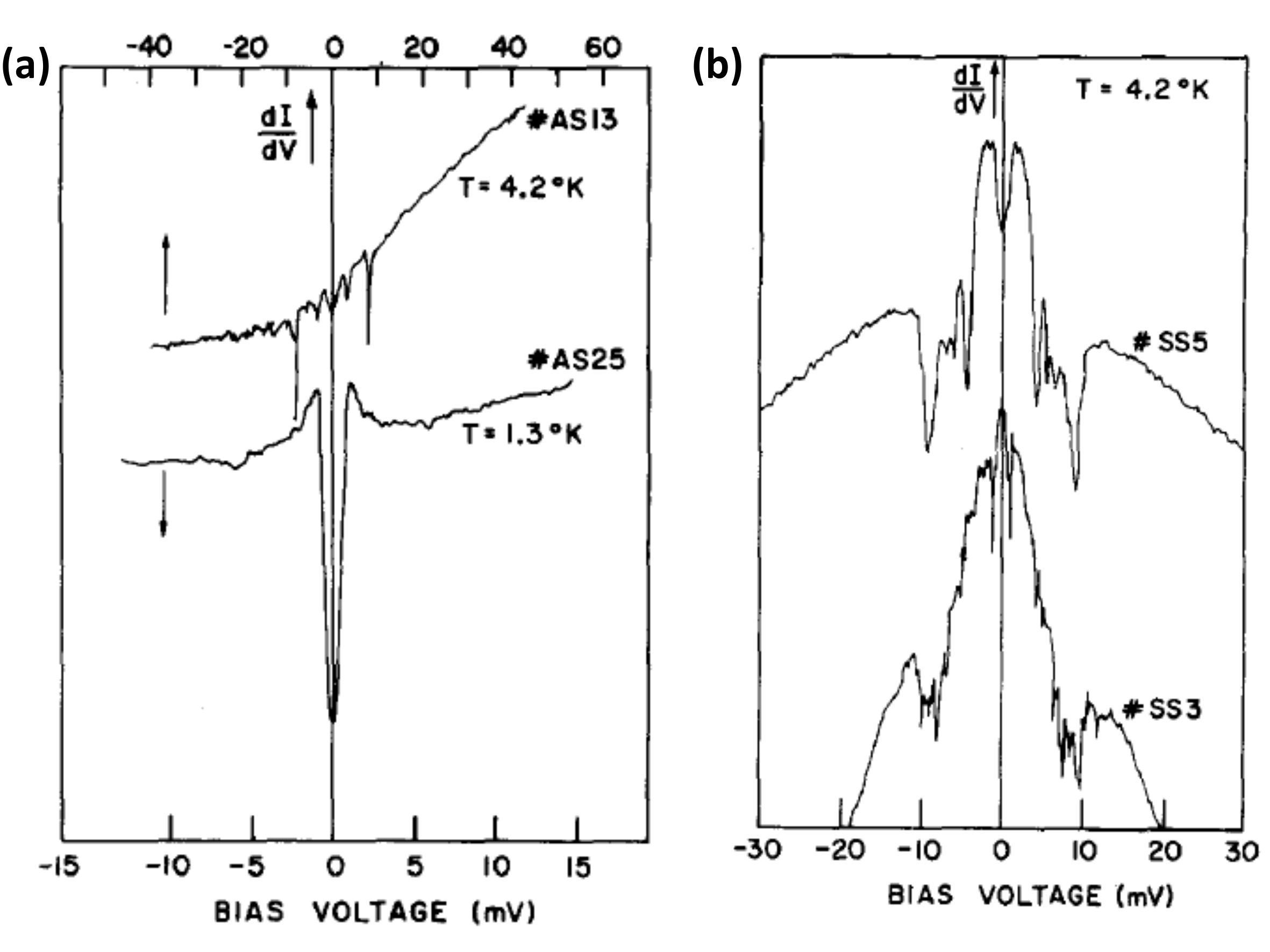}
	\caption{(a) Conductance spectra of Au-Si point contact junctions (b) Conductance spectra of silicon-silicon point contact junction.\cite{Tanaka}, reproduced with permission, 1974, Elsevier}
	\label{s1}
\end{figure}

\section*{VIII. Future outlook}
Superconducting devices have potential applications in various fields ranging from sensors and detectors to quantum information technology. However, all superconducting future advancements will require complex framework and the adopted technology must be perfectly managed in terms of reliability and reproducibility. In this quest, superconducting silicon can address the challenges owing to the established fabrication of high quality silicon based functional devices.
Superconductivity in silicon can provide an avenue towards designing superconducting nanomechanical resonators and Superconducting Quantum Interference Device (SQUID) on chip that could work as electro-magnetic sensors. 
Duvauchelle \textit{et al}. made an attempt to form SQUID from a monolayer superconducting silicon fabricated using the GILD technique.\cite{Duvauchelle} The device consists of two nano-sized bridges and in the fabricated device, magnetic flux could be modulated at low values of magnetic field and temperature. More practical devices can be implemented in future such as Josephson Field Effect Transistors comprising of superconducting drain and source contacts and current flowing through a channel modulated by an electrostatic gate. 
Josephson junction comprising of two superconducting materials with an insulating layer sandwiched between them, is the basic building block of most of the superconducting devices. Superconductor-metal-superconductor junction, an alternative to Josephson junctions, was proposed theoretically by De Gennes\cite{de Gennes} and experimentally investigated by Clarke.\cite{Clarke} When a normal metal and superconductor are electrically contacted, Cooper pairs can penetrate from the superconducting side to the normal metal, due to the proximity effect, a commonly known phenomenon in superconductivity.\cite{proximity} Researchers have studied a special category of this kind of junctions by replacing the metal by a degenerately doped semiconductor.\cite{T} The advantage of the superconductor-semiconductor-superconductor junctions lies in the modulation of the coupling strength between the superconductors by changing the doping/gate voltage and hence, carrier concentration in the semiconductor, which results in the variable supercurrent. In the case of semiconductor-superconductor junction, the Schottky barrier present at the semiconductor-superconductor interface limits the seamless flow of Cooper pairs from superconductor side to semiconductor side. In this respect, silicon(semiconductor)-silicon(superconductor) junctions can provide Schottky barrier free contacts. This can be made possible using the gas immersion laser doping (GILD) technique in which highly doped regions can be fabricated with abrupt interfaces. The ultimate control over supercurrent by varying the doping level in the semiconductor can lead to applications such as low power, ultrafast electronic devices.
Another important application of superconductivity in silicon can find in energy saving devices by reducing energy losses to a considerable extent. For modern day technology, there exists a huge requirement of various energy saving devices by limiting the energy losses to an extent with high efficiency. Especially, superconducting silicon can be a potential material for such devices including superconducting electric motors and generators, magnetic energy storage systems and so forth.\cite{Schneider}

\section*{IX. Conclusions} 

On account of the experimental and theoretical research efforts, the field of superconductivity in various structural forms of silicon has progressed enormously. Although advancement in structural phases prediction in silicon has identified a number of superconducting phases, an effective route towards discovering superconductivity in moderately doped silicon is required. On the other hand, discovery of superconductivity with high critical temperature is crucial if the promise of silicon as superconductor is to be realized for industrial purpose. Indeed, different structural forms of silicon have shown promising superconducting properties, but thorough experimental investigation is lacking. For example, nano-dimensional silicon quantum wells and Josephson junctions have presented the possibility of improved superconducting properties with relatively higher critical temperature. Silicene, the monolayer conterpart of silicon, can also serve as a promising candidate for miniaturized future superconductor technology due to strong electron-phonon coupling as predicted theoretically and verified experimentally with the possibility to achieve higher critical temperatures. For the fundamental understanding of superconductivity in silicon at nanoscale, point contact spectroscopy is an effective tool. Tip-induced superconductivity in silicon with different kinds of non-superconducting metallic probes can open a way towards more fundamental understanding of superconductivity in silicon with higher critical temperature. 
At last, an attempt is made to provide an overview of the developments and ongoing research in the field of superconductivity in silicon. The future research in this field will lead to advanced understanding of such exotic superconducting properties of silicon, especially towards high critical temperature to make this material more feasible for superconducting device applications.

\section*{Acknowledgements}
GS acknowledges the support of Swarnajayanti fellowship awarded by the Department of Science and Technology (DST), Govt. of India (grant number DST/SJF/PSA01/2015-16). MM acknowledges the support of NPDF-SERB fellowship (PDF/2020/002122).


\end{document}